\documentclass[%
aps,
prb,
showpacs,
reprint,
nofootinbib,
amsmath,amssymb,
]{revtex4-1}

\usepackage{graphicx}
\usepackage{dcolumn}
\usepackage{bm}

\begin{document}
\title{Three-body radiation dynamics in systems with anisotropic nanoparticles}
\author{Moladad Nikbakht}
\email{mnik@znu.ac.ir}
\affiliation{Department of Physics, University of Zanjan, Zanjan 45371-38791, Iran.}
\date{\today}

\begin{abstract}
The time evolution of temperatures of anisotropic nanoparticles in two and three-body systems are simulated for various relative orientations. Nanoparticles are immersed in a thermal bath at constant temperature. It is shown that in two-body systems, the relative orientation of nanoparticles could drastically affect the dynamics of temperature evolution and thermalization time scale. Moreover, in some configurations, the temperature difference in initial state has a minor effect on the dynamics of temperatures. In three-body systems, the orientation of the third nanoparticle influences the temperature dynamics, which allows one to control the thermalization time scales between anisotropic nanoparticles. Also, in addition to previously known contribution of the smallest distance between isotropic nanoparticles on the thermalization time scales, it is shown that the nanoparticles' orientations are more important in some particular arrangements.
\end{abstract}
\maketitle
\section{\label{sec1}Introduction}
It is well-known that the radiative heat transfer intensity can be sufficiently enhanced when the separation between objects is much smaller than the characteristic wavelength of thermal emission. This phenomenon is demonstrated experimentally for plane-plane and sphere-plane geometry and it is shown that the heat flow at nanoscale distances is greater than the black body radiation law posed by Stefan Boltzman \cite{DiMatteo1,Kittel4,rousseau2,Ottens3,Naray5,narayanaswamy6}. Several theories have been proposed to explain the heat transfer between simple particle geometries at nanoscale distances and have been widely used for two-body systems including parallel slabs \cite{ben22,Nefzaoui,Wang}, a particle in front of a planar surface \cite{rousseau2,biehs25,huth19},two spheres \cite{madrid8,saaih10,Nara11,madrid12,Manjavacas29}, and two anisotropic objects \cite{biehs20,Biehs23,incardone13}. There are series of theoretical studies in recent years concerning the many body effects in the radiative heat transfer \cite{golyk9,ben14,messina15,phan21,Simovski24,ben16}. It is shown in many publications that the heat flux exchange between two objects can be remarkably enhanced in a three-body system with respect to a two-body system, thanks to many-body interactions. In [\onlinecite{ben16}], Ban-Abdallah {\it et al}. used the fluctuational electrodynamics theorem to develop a theoretical formalism for radiative heat transfer between spherical nanoparticles in a many-body system in the dipole limit. They have shown a configurational resonance in heat exchange between nanoparticles due to many-body interactions. Similarly, the exaltation of the heat transfer in a three-body system of parallel slab is reported in [\onlinecite{Messina7}].
There are very few studies in the heat transfer between anisotropic materials. The effect of shape, orientation and polarizability tensor in the heat transfer between anisotropic objects allows for a large freedom of tunability. Early research on the heat exchange between anisotropic objects demonstrate that the relative orientation and polarizability tensors offer additional mechanism to tune the heat transfer, which is of interest for the design on new materials capable of tailoring heat propagation and applications in thermal sensors. Huth {\it et al}. studied the shape-dependence of heat transfer between a shperoidal nanoparticle and a planar surfaces for various distances \cite{huth19}. Compared to the sphere-plane geometry, they have shown that it is possible to control the heat transport in the system by the shape. Biehs {\it et al}. showed that the heat exchange between two anisotropic media can be significantly larger than those traditionally measured between two isotropic materials \cite{biehs20}. The strong dependence of heat exchange between two anisotropic nanoparticles on their shapes and relative orientations has been recently reported and interpreted as a heat transfer switch \cite{incardone13}. In our recent work \cite{Nikbakht28}, we used the fluctuational electrodynamics to introduce a formalism to take into account the change in particle's shape and more generally the polarizability tensors in a many-body system. We showed the heat exchange in a system of anisotropic nanoparticles can be tuned drastically depending on relative orientations and polarizability tensors.

Most of the theoretical efforts in the investigation of radiative heat transfer have been restricted to the steady state regime. The issue of radiative thermalization and dynamics of temperatures has been addressed in few papers and very recently some concrete ideas have been put forward. Tschkin {\it et al}. have investigated the radiative cooling of a spherical nanoparticle close to an interface with fixed temperature \cite{tschikin27}. They have shown that the thermal relaxation time of the nanoparticle varies rapidly for distances smaller than the thermal wavelength and is very sensitive to the initial temperature differences. Dyakov, {\it et al}. studied the propagation of thermal radiation in a two layered structures. In [\onlinecite{dyakov18}], they assumed that the temperature of one of the plates is changed due to the radiative heat transfer from another plate at constant temperature which acts as a heat source. They have shown that the thermalization time scale between the plates strongly depends on the separation distance. The dynamics of heat transfer between isotropic nanoparticles is well treated by Messina and his co-workers with consideration of many body effects \cite{messina26}. Using a purely fluctuational electrodynamics approach, they studied the time evolution of temperatures of spherical nanoparticles with respect to several geometrical arrangements. It was suggested that many-body interactions might lead to rapid thermalization. They concluded that the smallest distance between nanoparticles in a system determines the time scale on which the heat flux is exchanged between the particles. 

In this paper, we make a step forward in simulating the dynamics of heat transfer between anisotropic nanoparticles which are immersed inside a thermal bath. To do this, we use the fluctuational electrodynamics theorem which relates the power spectral density of fluctuating charge density to the local temperature and frequency dependent relative dielectric response of nanoparticle \cite{book34}. Using this theorem and properties of the dyadic Green's function as well as polarizability tensors, allows studying the phenomenon of heat transfer between anisotropic nanoparticles at different temperatures. For the sake of simplicity, we will restrict ourselves to nanoparticles with spheroidal shape only which made of silicon carbid (SiC). We are interested in the effects of anisotropy, i.e., the shape and the orientation of shperoidal nanoparticles, on the dynamics of temperatures as well as the thermalization time scales in two and three-body systems. We explain our findings by studying the heat transfer as a function of relative orientation of nanoparticles.

The structure of the paper is as follows. Our study of dynamics of temperatures in system of anisotropic nanoparticles begins in section.~\ref{sec2}. But before digging in, we will discuss briefly the basic ideas of heat exchange between anisotropic nanoparticles in presence of an external thermal bath. In Sec.~\ref{sec3} we consider dynamics of heat transfer between two spheroidal nanoparticles of silicon carbid (SiC) immersed in a thermal bath. We studied the dependence of the temperatures' dynamics on the relative orientation of nanoparticles. To give an intuitive realization, the phase portraits of the system are also presented for various configurations. Section~\ref{sec4} contains the results of temperature dynamics in three-body system of anisotropic nanoparticles. This section is divided into four subsections which the dependences of thermalization time scales on the orientation of the third particle are discussed in Secs.~\ref{sec4A} to  \ref{sec4D}. Finally, we draw our conclusion in Sec.~\ref{sec5}.

\section{\label{sec2} dynamics study of heat exchange between anisotropic particles}
Before we go through the dynamics of heat transfer between anisotropic particles, let us focus on the problem of radiative heat exchange in a system of anisotropic nanoparticles without considering the temporal evolution of temperatures. Our theoretical framework for heat exchange between anisotropic nanoparticles has been described in previous publication [\onlinecite{Nikbakht28}] and is briefly summarized here. We considered an energy exchange problem between $N$ anisotropic nanoparticles identified for convenience by indexes $i=1, 2, \cdots, N$ with positions (${\bf r}_i$), polarizability tensors (${\hat {\bm \alpha}_i}$), and temperatures ($T_i$) . In order to take into account the radiation impinging on such a system from the outside, we assumed that this collection is immersed in a thermal bath at a constant temperature ($T_b$). 
Based on the assumption that nanoparticles behave as fluctuating dipoles at different temperatures, the fluctuation-dissipation theorem provides the energy which is dissipated in each nanoparticles. The nanoparticles themselves interact through the radiation fields and on the other side with the thermal bath and the total radiation power dissipated in each particle were obtained as:
\begin{equation}
\mathcal{P}_i={\mathcal F}_i(T_i)+\sum_{j \neq i}{\mathcal F}_{i,j}(T_j)+\sum_{jj'}{\mathcal F}_{i,jj'}^{b}(T_b)
\label{eq1}
\end{equation}
with
\begin{subequations}
\begin{eqnarray}
&&{\mathcal F}_i(T_i)={\tt Im}{\int}_0^\infty\frac{d\omega}{\pi}{\mathbb Tr}[{\hat{\bf A}}_{ii}{\tt Im}({\hat {\bm \chi}}_{i}){\hat{\bf C}}_{ii}^\dag]\Theta(\omega,T_i),~~\\
&&{\mathcal F}_{i,j}(T_j)={\tt Im}\int_0^\infty\frac{d\omega}{\pi}{\mathbb Tr}[{\hat{\bf A}}_{ij}{\tt Im}({\hat {\bm \chi}}_{j}){\hat{\bf C}}_{ij}^\dag]\Theta(\omega,T_j),~~\\
&&{\mathcal F}_{i,jj'}^{b}(T_b)={\tt Im}\int_0^\infty\frac{d\omega}{\pi}{\mathbb Tr}[{\hat{\bf B}}_{ij}{\tt Im}({\hat {\bm G}}_{jj'}){\hat{\bf D}}_{ij'}^\dag]\Theta(\omega,T_b),~~~~~~~
\end{eqnarray}
\label{eq2}
\end{subequations}
\\Where ${\mathcal F}_{i}(T_i)$ denotes the so-called self-emission by {\it i}-th particle corresponding to the heat lost by its radiation in presence of other particles in the system, ${\mathcal F}_{i,j\neq i}(T_j)$ is the radiative heating of the {\it i}-th particle due to radiation of the {\it j}-th one, and finally ${\mathcal F}_{i,jj'}^{b}(T_b)$ is a portion of power which is absorbed in the {\it i}-th particle due to the radiation of the thermal bath. In these equations, $\Theta(\omega,T)=\hbar\omega[1+2n(\omega,T)]/2$ and $n(\omega,T)$ is the Bose-Einstein energy distribution function and we have introduced ${\hat {\bm \chi}}={\hat {\bm \alpha}}+k^2{\hat {\bm \alpha}}{\hat {\bm G}}_0^\dag{\hat {\bm \alpha}}^\dag$. Here ${\hat{\bf G}}_{ij}={\hat{\bf G}}_{ji}$ are the electric Dyadic Green functions, ${\hat{\bf G}}_{0}=i\frac{k}{6\pi}{\hat{\bf 1}}$, and $ K=\omega/c$. The physical properties of nanoparticles and the mathematical effects of many-body interactions are contained in the coupling matrixes ${\hat{\bf A}}$, ${\hat{\bf B}}$, ${\hat{\bf C}}$ and ${\hat{\bf D}}$. For the special case of three anisotropic nanoparticles, we find
\begin{subequations}
\begin{eqnarray}
&&{\hat{\bf A}}=
\begin{bmatrix}
{\hat{\bf 1}} & {-k^2{\hat{\bm\alpha}}_1\hat{\bf G}}_{12} & {-k^2{\hat{\bm\alpha}}_1\hat{\bf G}}_{13}\\ 
{-k^2{\hat{\bm\alpha}}_2\hat{\bf G}}_{21} & {\hat{\bf 1}}& {-k^2{\hat{\bm\alpha}}_2\hat{\bf G}}_{23}\\
{-k^2{\hat{\bm\alpha}}_3\hat{\bf G}}_{31}& {-k^2{\hat{\bm\alpha}}_3\hat{\bf G}}_{32} &{\hat{\bf 1}}
\end{bmatrix}^{-1},~~\\
&&{\hat{\bf B}}=k^2{\hat{\bf A}}
\begin{bmatrix}
{\hat{\bm\alpha}}_1 & {\hat{\bf 0}} & {\hat{\bf 0}}\\ 
{\hat{\bf 0}} & {\hat{\bm\alpha}}_2& {\hat{\bf 0}}\\
{\hat{\bf 0}}&{\hat{\bf 0}}&{\hat{\bm\alpha}}_3
\end{bmatrix},~~\\
&&{\hat{\bf C}}=k^2
\begin{bmatrix}
{\hat{\bf G}}_{0}&{\hat{\bf G}}_{12} & {\hat{\bf G}}_{13}\\ 
{\hat{\bf G}}_{21} & {\hat{\bf G}}_{0}& {\hat{\bf G}}_{23}\\
{\hat{\bf G}}_{31}&{\hat{\bf G}}_{32}&{\hat{\bf G}}_{0}
\end{bmatrix} {\hat{\bf A}},~~\\
&&{\hat{\bf D}}={\hat{\bm 1}}+
\begin{bmatrix}
{\hat{\bf G}}_{0}&{\hat{\bf G}}_{12} & {\hat{\bf G}}_{13}\\ 
{\hat{\bf G}}_{21} & {\hat{\bf G}}_{0}& {\hat{\bf G}}_{23}\\
{\hat{\bf G}}_{31}&{\hat{\bf G}}_{32}&{\hat{\bf G}}_{0}
\end{bmatrix}{\hat{\bf B}}.~~
\end{eqnarray}
\label{eq3}
\end{subequations}
The frequency-dependent relative dielectric permittivity of SiC is well described by lorentz model \cite{karl35}:
\begin{equation}
\varepsilon(\omega)=\varepsilon_\infty\bigg(1+\frac{\omega_L^2-\omega_T^2}{\omega_T^2-\omega^2-i\Gamma\omega}\bigg)
\label{eq4}
\end{equation}
with $\omega_L=969$ cm$^{-1}$, $\omega_T=793$ cm$^{-1}$, $\Gamma=4.76$ cm$^{-1}$ and $\varepsilon_\infty=6.7$. This relative permittivity is used to calculate the dressed polarizability tensors of nanoparticles. In the case of small isotropic non-spherical particles (or different extensions, including anisotropic spherical particles or anisotropic non-spherical particles), the dressed polarizability tensor can be expressed as \cite{albaladejo31}:
\begin{equation}
{\hat{\bm \alpha}}_i={\hat{\bm \alpha}}_h\bigg\{{\hat{\bf 1}-i\frac{k^3}{6\pi}{\hat{\bm\alpha}}_h}\bigg\}^{-1},\label{eq5}
\end{equation}
with
\begin{equation}
{\hat{\bm\alpha}}_h=v_i({{\hat{\bm \varepsilon}}}-{\hat{\bf 1}})\bigg\{{{\hat{\bf 1}}}+({\hat{\bf L}}-k^2{\hat{\bf M}})({{\hat{\bm \varepsilon}}}-{\hat{\bf 1}})\bigg\}^{-1},\label{eq6}
\end{equation}
where ${\hat{\bf L}}$ and ${\hat{\bf M}}$ are the singular and non-singular part of the depolarization dyadic \cite{book33}.
For special case of small prolate spheroidal nanoparticles, the polarizability tensor is diagonal in the principal-axis system of each particle and the components of the polarizability tensor along and across the rotation axis of a spheroidal nanoparticle are given by:
\begin{equation}
{\hat\alpha}_{h,\beta}=\frac{4}{3}\pi R_bR_s^2\frac{\varepsilon (\omega)-1}{1+L_{\beta}[\varepsilon(\omega)-1]}, (\beta=b,s),
\label{eq7}
\end{equation}
where $R_b$ and $R_s$ are the major and minor axeses of spheroid and the depolarization factors take the form \cite{book36}:
\begin{subequations}
\begin{eqnarray}
{L}_s&=&\frac{1}{2}(1-L_b),\label{eq8a}\\
{L}_b&=&\frac{1-e^2}{2e^3}\bigg[ ln\bigg(\frac{1+e}{1-e}\bigg)-2e\bigg],\label{eq8b}
\end{eqnarray}
\end{subequations}
with 
\begin{equation}
e^2=1-\frac{R_s^2}{R_b^2}.
\label{eq9}
\end{equation}

Now that we have established the formalism of heat exchange between anisotropic nanoparticles, we can present a framework for dynamics of temperatures. The time evolution of the temperatures in system is a solution of the nonlinear coupled system of differential equation:
\begin{equation}
\rho_ic_iv_i\frac{\partial {T_i}}{\partial{t}}=\mathcal{P}_i(T_1, T_2,\cdots,T_N,T_b)~~~~~~~(i=1, 2, \cdots, N).
\label{eq10}
\end{equation}
where $v_i$ is the volume of the particle with mass density $\rho_i$ and heat capacity $c_i$. The temperature-dependent heat capacity of SiC is calculated as follows \cite{book37}:
\begin{equation}
c=1.267+0.049\times 10^{-3}T-1.227\times 10^{5}T^{-2}+0.205\times 10^{8}T^{-3}
\label{eq11}
\end{equation}
We note that the right-hand side of Eq.~(\ref{eq10}) has N nonlinear terms with a strong nonlinearity in temperatures. We need the initial values of temperatures as well as the bath's temperature to determine the solution uniquely. The solutions of this equation could be visualized as trajectories flowing through an N-dimensional phase space with coordinates ($T_1,\cdots,T_N$). We will discuss the dynamics of temperatures, and will show that the qualitative structure of the temperatures' dynamics can change as the relative orientation of particles is varied.

\section{\label{sec3}Dynamics of heat transfer in two-body system}

In this section we study the temperature evolution in two-body systems and visualize the dependence of temperature dynamics on the nanoparticles relative orientation. The system consists of two prolate spheroidal nanoparticles (major to minor axis ratio, $\eta=4$) with a fixed center to center separation, $d=800$ nm. The particles are heated up to initial temperatures $T_1(0)$ and $T_2(0)$ and immersed in a thermal bath at fixed temperature $T_b$. The nanoparticles themselves interact with each other and also with the thermal bath. Our aim is to calculate the particles' temperatures, $T_1$ and $T_2$, as a function of time. So, Eq.~(1) is used to calculate the total power dissipated in each particle and the dynamics of temperatures are calculated according to the energy balance equation, Eq.~(\ref{eq10}). Since, the thermalization rate in each particle due to thermal conduction by heat carriers is much larger than the heat transfer rate between the particle and the rest of the system (including other particles and thermal bath) via radiation, it ensures that the energy fluxes supported by heat carrier inside each particle is large enough to treat the temperatures as homogeneous inside each particle during the dynamical evolution of the system.

\begin{figure}[t]
\includegraphics[scale=1.1]{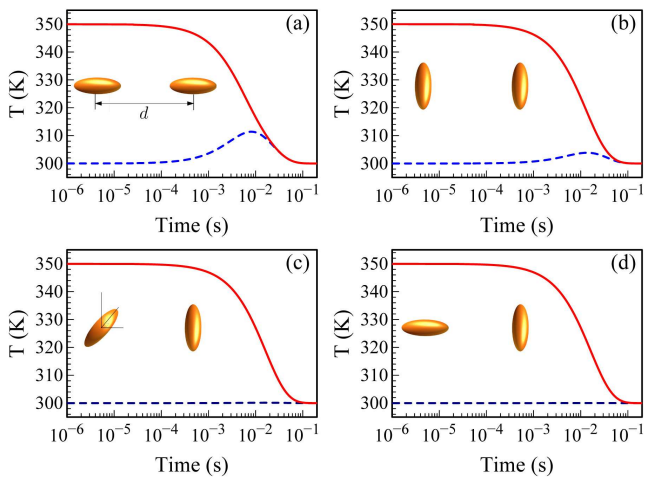}
\caption{(Color online) Time evolution of the temperatures in a two-body system for (a) tip-tip, (b) side-side, (c) cross and (d) tip-side (side-tip) configurations. The distance between particles is $d=800$ nm and initial temperatures are $T_1=350$ K, $T_2=300$ K and thermal bath is at fixed temperature $T_b=300$ K. Nanoparticles are prolate spheroid with major to minor aspect ratio $\eta=R_b/R_s=4$. The red (solid) line corresponds to the hotter particle while the blue (dashed) line is for cold one. }
\label{fig1}
\end{figure}
In order to show the relative orientation effect on the dynamics of heat transfer, particles are arranged in the system with one of the main arrangements, namely the tip-tip, side-side, tip-side (side-tip), and cross configurations. Figure~(\ref{fig1}) shows solution to Eq.~(\ref{eq10}) for these types of configurations. The initial temperatures are $T_1=350$ K and $T_2=300$ K and particles are separated $d=800$ nm apart in a thermal bath at $T_b=300$ K. The fixed point of Eq.~(\ref{eq10}) corresponds to the thermal equilibrium state of the system. In the presence of thermal bath (at constant temperature) and absence of any other external sources, the particles would thermalize to the temperature of the bath. On the other hand, there is a fixed point at bath's temperature, regardless of the nanoparticles' characteristics or initial temperatures. We would expect no change in temperatures in thermal equilibrium and from Fig.~(\ref{fig1}) it is clear that it takes approximately $0.1$ second for particles to thermalize with bath ( i.e., $T_{1,2}\rightarrow 300$ K) which is independent from the relative orientation of the nanoparticles. However, it can be seen that there exist two types of time scales in the curves; one corresponds to the thermalization between two particles (smaller time scale) and the other for thermalization of particles with the bath (larger time scale). The former is mainly due to the intra-particle heat exchange while the later is caused by the radiative heating of nanoparticles by the thermal bath.

In the case of tip-tip configuration, the first thermalization is $\sim$ 1 order of magnitude faster than the second thermalization. For the side-side configuration, the early thermalization time scale increased slightly and the first thermalization temperature tends to $T_b$. This is because the coupling between particles in this configuration is small in comparison with the tip-tip case and the heat transported from the particle at larger temperature to the colder one could slightly compensate the decrease of the temperature of the colder one due to its thermal emission. With decrease of ordering in the system, by going from tip-tip or side-side to cross or tip-side configurations, the heat exchange between particles drops several orders of magnitude \cite{Nikbakht28}. As a result, the rate of energy lost in the cooler particle is more than received from the hotter one and it seems that particles are thermalize by the bath independently.
\begin{figure}[t]
\includegraphics[scale=1]{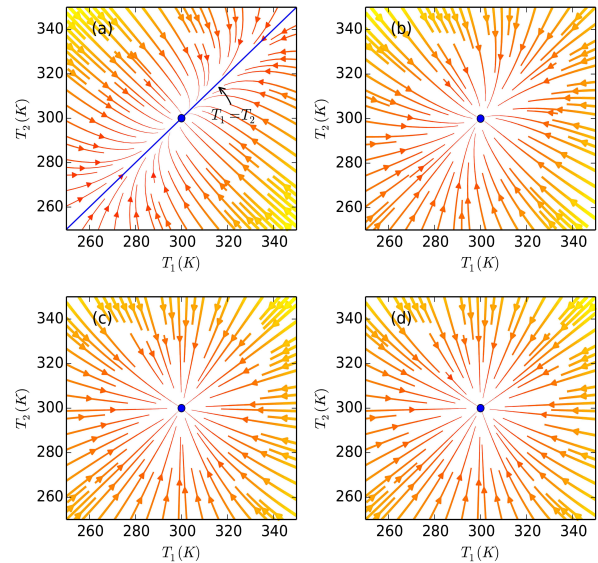}
\caption{\label{fig2} Phase portrait of the two-body system immersed in a thermal bath at temperature $T_b$=300 K for (a) tip-tip, (b) side-side, (c) cross and (d) tip-side configurations. The thickness of the trajectories show the thermalization rate.}
\end{figure}

The phase portraits for different configurations of particles in the two-body system are shown in Fig.~(\ref{fig2}). Each of these configurations has a stable fixed point at $T^*=T_b$ (Solid blue dot) and all trajectories approach it as $t\rightarrow \infty$. However, the direction of approach depends on the initial temperature difference as well as the relative orientation of the particles. Moreover, the thickness of the trajectories shows the thermalization rate which is faster for large temperature differences and the phase point slows down as reaches the proximity of the thermalization point. 

In the case of tip-tip configuration, as shown in Fig.~(\ref{fig2}a), all trajectories slam straight onto the {\it slower direction} (here, the $T_1=T_2$ direction) which corresponds to intra-particle thermalization regime, then slowly ooze along this curve until they reach a fixed point. The point at which the trajectory reaches this line is the intra-particles' thermalization temperature. The approach to the fixed point is tangent to the $T_1=T_2$ direction which the thermalization by bath occurs. On the other hand, if we look backwards along a trajectory ($t\rightarrow -\infty$), then the trajectories all become parallel to the {\it faster decaying direction} (here, the $T_1+T_2-2T_b=0$ direction). 

As the relative orientation changes to side-side configuration, the coupling between the nanoparticles decreases an order of magnitude in comparison to tip-tip configuration \cite{Nikbakht28}. As a result, the time scale for thermalization of the particles with each other increases and the decay to the slower direction decreased. The trajectories shows faster decay along the $T_1=T_2$ direction than previous configuration and approaches the fixed point monotonically as $t\rightarrow \infty$, [see Fig.~(\ref{fig2}b)]. Furthermore, similar to the tip-tip configuration, intra-particle thermalization temperatures depend strongly on the initial temperatures of the particles.

In the case of cross and tip-side configurations, Figs.~(\ref{fig2}c) and (\ref{fig2}d) respectively, the coupling between nanoparticles is very week in comparison to the previous configurations. In these cases, the thermalization by thermal bath is more effective and $T^*=T_b$ is a symmetrical node or star shaped fixed point of the system which represents an attractor that the flows are toward it. Since the decay rates of temperatures are approximately the same, all trajectories are approximately straight lines through the fixed point. Moreover, the time it takes for particles 1 and 2 to thermalize with each other does not appreciably depends on their initial temperatures and is of order of thermalization with bath. From these phase portraits it becomes apparent that both the temperature differences and relative orientation of the particles determine the time scale on which the heat flux is exchanged between anisotropic nanoparticles.

\section{\label{sec4}Dynamics of heat transfer in three-body systems}
Let us now calculate the time evolution of the temperatures in a three-body system of anisotropic nanoparticles. For the sake of simplicity, consider a two-body system at which particles 1 and 2 are oriented in one of the configuration discussed in previous section. Now, a third nanoparticle labeled with $i=3$ with the same physical properties, is placed between nanoparticles 1 and 2 [see for example Fig.~(\ref{fig4}a)]. In such a system, each of the nanoparticles exchanges heat with the two others and the bath. Once again the dynamics of temperatures obey Eq.~(\ref{eq10}) and we will show that the system undergoes different thermal dynamics depending on the orientation of the third particle. Intuitively we know that the coupling between 1 and 2 affects by the presence of the third particle and leads to an enhancement (reduction) of the radiative heat exchange, therefore, to shortening (enlargement) of time scale of intra-particle thermalization. Moreover, addition of a new particle would alter the coupling between each particle and the thermal bath and apart from the intra-particle interaction it would affect the thermalization time scales. From a dynamical point of view, addition of the third particle increases the phase space dimension which might allows for more possible dynamical properties and we may encounter some new phenomena. As an example, adding a particle introduces new intra-particle thermalization time scales to the problem which depends on initial temperatures.

To compare the results with the previous section, we assume that particles 1 and 2 are arranged in either tip-tip, side-side, cross or tip-side configuration and the evolution of the temperatures are discussed for a particular orientation of the third particle in each of these types. The time evolution of the temperatures are calculated for the initial temperatures $T_1=350$ K and $T_2=T_3=300$ K, and the thermal bath is maintained at fixed temperature $T_b=300$ K. 

\begin{figure}[t]
\includegraphics[scale=.8]{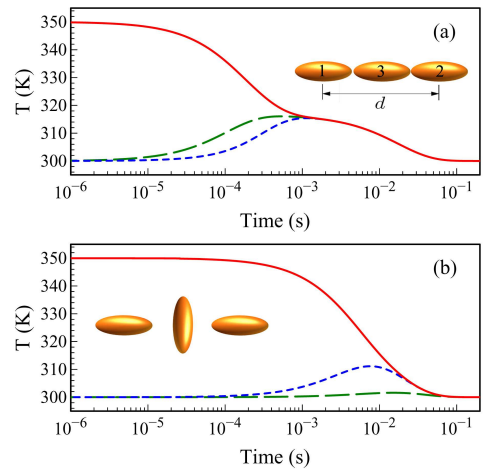}
\caption{\label{fig3} (Color online) Dynamics of temperatures in a three-body system. Particles 1 and 2 are oriented in a tip-tip configuration and a third particle is placed between them to form a (a) tip-tip-tip and (b) tip-side-tip configuration. The separation between 1 and 2 is $d=800$ nm as the two body case and initial temperatures are $T_1=350$ K, $T_2=300$ K and $T_3=300$ K, respectively. The collection is immersed in a thermal bath at $T_b=300$ K. The red (solid) line corresponds to the hotter particle (the first particle), the blue (dashed) line is for cold one (the second particle) and the green (long-dashed) line is for the third particle.}
\end{figure}

\subsection{\label{sec4A} Three-body system: particles 1 and 2 in tip-tip configuration}

Let us first consider the thermal evolution problem in a system where particles 1 and 2 form a tip-tip configuration. Figures \ref{fig3}a and \ref{fig3}b represents the time evolution of the temperatures calculated for two particular arrangements, namely, tip-tip-tip and tip-side-tip, respectively. 

In Fig.~(\ref{fig3}a), nanoparticles are aligned along the longer spheroid axis and the intra-particle couplings are very strong. Surprisingly, it can be seen that the presence of the third particle reduces the time scale for thermalization between 1 and 2 of about an order of magnitude in comparison with the two-body system. Moreover, the heat exchange between 1 and 3 is enough to let the third particle thermalize by first one even slightly faster than the time scale of thermalization of 1 and 2. This is explained by the fact that the coupling between 1 and 3 is larger than the coupling between 1 and 2 due to separation distances. After such an intra-particle thermalization, it takes a long time for the phase point ($T_1=T_2=T_3$) to pass through such a {\it bottleneck} state before making a transition to the final thermalization state of the system. Since the coupling with the thermal bath does not change significantly by the presence of the third particle, the time it takes for the system to reach the steady state is as the same order in the two-body system.

As the third particle rotates to form a tip-side-tip configuration, Fig.~(\ref{fig3}b), the coupling between 1 and 2 drops an order of magnitude in comparison with the previous case \cite{Nikbakht28}. The tip-side coupling between 3 and 1 (or 2) is not large enough and it has a trivial role in the dynamics of temperatures of particle 1 and 2. As a result, the temperature evolutions of particles 1 and 2 are similar to the two-body case [see Fig.~(\ref{fig1}b)]. Moreover, particle 3 is not affected by particle 1, as much as particle 2, even though it is closer to it and has the same initial temperature as particle 2. Basically, this corresponds to the nature of couplings between anisotropic nanoparticles. On the other hand, while the smallest distances between particles determines the thermalization time scales in systems with isotropic particles \cite{messina26}, the orientations (or more generally, the polarizability tensors) play a crucial role when we are dealing with anisotropic particles.

\begin{figure}[t]
\includegraphics[scale=.8]{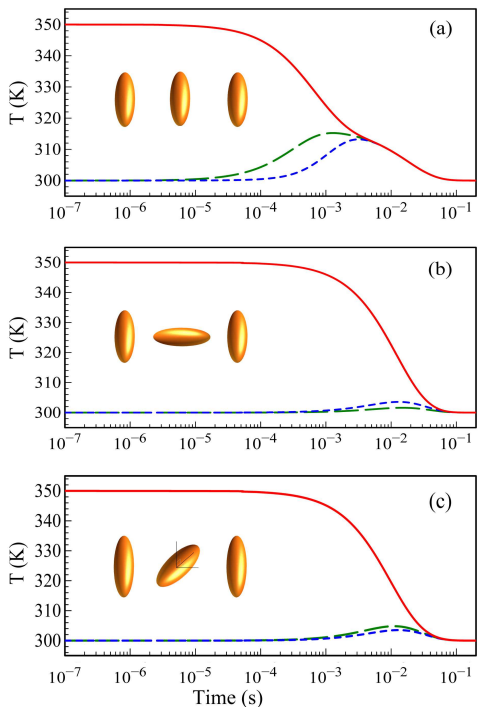}
\caption{\label{fig4} (Color online) Dynamics of temperatures in a three-body system. Particles 1 and 2 are oriented in a side-side configuration and a third particle placed between them to form a (a) side-side-side, (b) side-tip-side, and (c) cross-cross configuration. }
\end{figure}

\begin{figure}[t]
\includegraphics[scale=.8]{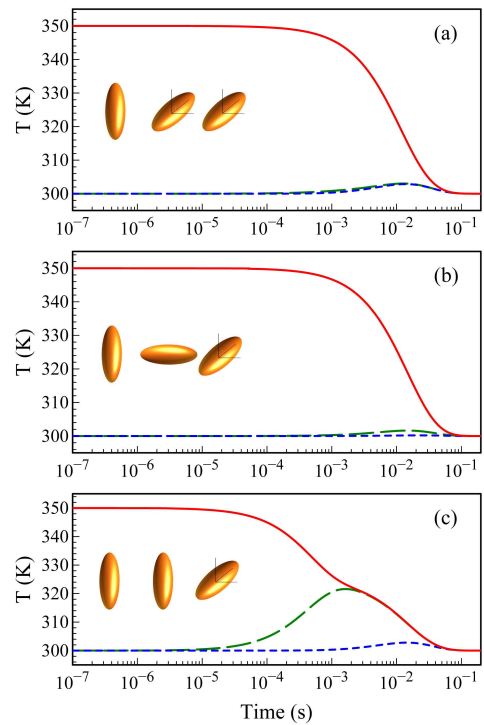}
\caption{\label{fig5} (Color online) Dynamics of temperatures in a three-body system. Particles 1 and 2 are oriented in a cross like configuration and a third particle placed between them to form a (a) cross-side-side, (b) side-tip-side, and (c) side-side-cross configuration. }
\end{figure}
\subsection{\label{sec4B} Three-body system: particles 1 and 2 in side-side configuration}
There exist three particular orientations in adding a third particle to a two-body system with side-side configuration. Figure (\ref{fig4}a) shows the temperature evolution for side-side-side configuration. Since the relative orientation of each couple of particles in the system are identical (side-side form), the intra-particle distances determines the time scales on which the heat flux is exchanged between the particles. In this case, particle 3 is closer to particle 1 and as a result it is heated faster than particle 2. Once again, the shift in the timescales can be attributed to the presence of the third particle which acts as a bridge for indirect heat transfer between particles 1 to 2. 

The results of the temperature evolution of particles for other orientation of the third particle are shown in Figs.~(\ref{fig4}b) and (\ref{fig4}c), for side-tip-side and cross-cross configurations, respectively. The orientation of the third particle does not affect the intra-particle coupling between particles 1 and 2 significantly. Moreover, particle 3 is heated slightly more by particle 1 in the cross-cross configuration before the final thermalization with the thermal bath.

\subsection{\label{sec4C} Three-body system: particles 1 and 2 in cross configuration}
The results for dynamical evolution of temperatures in a three-body system in which particles 1 and 2 oriented in cross like form are illustrated in Figs.~(\ref{fig5}a)-(\ref{fig5}c). As can be seen in Fig.~(\ref{fig5}a), the stronger coupling between particle 3 and 2 results in a similar thermal evolution of these particles toward the steady state of the system. In figure.~\ref{fig5}b, the tip-side coupling of particle 3 with 1 (and 2) leaves the dynamics of temperatures unchanged. On the contrary, as we can see in Fig.~(\ref{fig5}c), the coupling between particles 3 and 1 is more effective and results in a faster thermalization between them. This coupling causes a drastic change in the thermal evolution of the first particle.

\begin{figure}[t]
\includegraphics[scale=1.1]{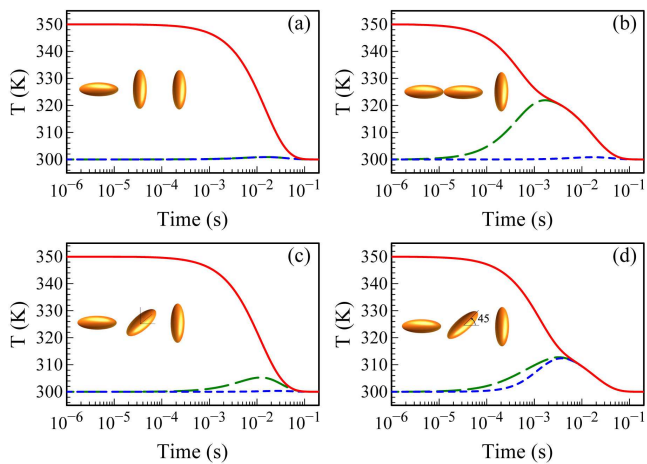}
\caption{\label{fig6} (Color online) Dynamics of temperatures in a three-body system. Particles 1 and 2 are oriented in a tip-side configuration and a third particle placed between them to form a (a) tip-side-side, (b) tip-tip-side, (c) tip-side-cross, and (d) tip-oblique-side configuration.}
\end{figure}

\subsection{\label{sec4D} Three-body system: particles 1 and 2 in tip-side configuration}
Now, consider a system in which particles 1 and 2 are arranged in a tip-side configuration. By adding a third particle, with one of the three mentioned orientations, three possible configurations are formed [see Figs.~(\ref{fig6}a)-(\ref{fig6}c)]. It can be seen in these figures that the temperature evolution of particle 2 is not influenced by the presence of particle 3. However the intra-particle thermalization between particles 1 and 3 is very sensitive to the orientation of the third particle and specially pronounced in a tip-tip-side configuration shown in Fig.~(\ref{fig6}b). 
The more interesting case is for configuration in which the third particle rotates by angle 45 degree in the plane of particles 1 and 2, namely tip-oblique-side configuration as shown in Fig.~(\ref{fig6}d). Upon inspection of low coupling between particles 1 and 2 observed in previous configurations, we find that the temperature evolution in tip-oblique-side is largely decided by three-body interactions. Here, the third particle appears to remove energy from the first particle and re-inject almost all of it to the second one. One can see that the temperature evolutions are drastically affected by the orientation of the third particle and the thermalization between particles happens very fast.

\section{\label{sec5}conclusion}
Using the method of fluctuational electrodynamics in conjunction with the properties of the dyadic Green's function as well as polarizability tensors, we have investigated the dynamics of temperatures in two (and three) body systems of anisotropic nanoparticles. Nanoparticles are immersed in a thermal bath at constant temperature and the dynamics of temperatures are simulated with respect to the relative orientation of nanoparticles. In the case of two-body system we showed the drastic effect of the relative orientation of nanoparticles on the dynamics of temperatures in the phase space of the system. In the case of three-body systems, we have found that the orientation of the third nanoparticle influences the dynamics of temperatures in comparison to the two-body systems. Also, in addition to the aforementioned works on the sensitivity of the heat transfer on the separation distances, it turns out that the dependence of the heat transfer on the shape and relative orientation of particles can be much larger than the smallest distances in some cases. Although we have considered the heat transfer dynamics for spheroidal particles, our study reveals the hidden characteristics of heat transfer between arbitrary shaped (and/or polarizability tensor) objects. We believe that our findings are essential for understanding the heat transfer in anisotropic systems and the physical results of this work relevant to control the thermalization time scales in nanoscale devices.

\bibliography{manuscript}

\end{document}